\documentclass[notitlepage,superscriptaddress]{revtex4}

\usepackage{amsmath,amssymb}
\usepackage{latexsym}
\usepackage[bookmarks=false,colorlinks=true]{hyperref}
\usepackage{bm}
\usepackage{color}

\numberwithin{equation}{section}

\newcommand{\ba}{\begin{array}}
\newcommand{\ea}{\end{array}}
\newcommand{\be}{\begin{equation}}
\newcommand{\ee}{\end{equation}}
\newcommand{\bea}{\begin{eqnarray}}
\newcommand{\eea}{\end{eqnarray}}

\begin{document}
\begin{flushright}
ITP--UH--6/16
\end{flushright}

\title{Spherical Calogero model with oscillator/Coulomb potential: quantum case}

\author{Francisco Correa}
\email{francisco.correa@itp.uni-hannover.de}
\affiliation{Leibniz Universit\"at Hannover, Appelstrasse 2, 30167 Hannover, Germany}
\author{Tigran Hakobyan}
\email{tigran.hakobyan@ysu.am}
\affiliation{Yerevan State University, 1 Alex Manoogian Street, Yerevan, 0025, Armenia}
\affiliation{Tomsk Polytechnic University, Lenin Avenue 30, 634050 Tomsk, Russia}
\author{Olaf Lechtenfeld}
\email{lechtenf@itp.uni-hannover.de}
\affiliation{Leibniz Universit\"at Hannover, Appelstrasse 2, 30167 Hannover, Germany}
\author{Armen Nersessian}
\email{arnerses@ysu.am}
\affiliation{Yerevan State University, 1 Alex Manoogian Street, Yerevan, 0025, Armenia}
\affiliation{Tomsk Polytechnic University, Lenin Avenue 30, 634050 Tomsk, Russia}

\begin{abstract}
We consider the quantum mechanics of Calogero models in an oscillator or Coulomb potential
on the $N$-dimensional sphere. Their Hamiltonians are obtained by an appropriate Dunkl deformation
of the oscillator/Coulomb system on the sphere and its restriction to (Coxeter reflection) symmetric
wave functions. By the same method we also find the symmetry generators and compute their algebras.

\end{abstract}

\maketitle

\section{Introduction}
\noindent
The rational Calogero model~\cite{calogero} and its various generalizations, based on arbitrary Coxeter
root systems~\cite{calogero-root}, continues to attract much interest due to the rich internal structure and
numerous applications.
In its simplest incarnation, for $A_{N-1}$ roots system,
it describes $N$ particles on a line with a pairwise inverse-square interaction potential.
An external oscillator potential preserves its integrability~\cite{calogero,moser}.
Moreover, these models were found to be superintegrable,
i.e.\ possessing $2N{-}1$ functionally independent constants of motion~\cite{woj83}.

There are two powerful tools for the study of Calogero models: the matrix-model approach~\cite{kazhdan}
and the exchange~operator (or Dunkl~operator) formalism~\cite{poly92,brink}.
For reviews on the subject see~\cite{calogero-review,calogero-root}.
In the  matrix-model approach, one starts from a free-particle system on the space of $N\times N$ Hermitian matrices,
so that the (discrete) permutation symmetries are absorbed in the natural SU$(N)$ invariance.
The SU$(N)$ reduction of this system in the minimal gauge yields the original $N$-particle Calogero model.
In the operator-exchange formalism, the Calogero interaction is generated by a Dunkl deformation of the momenta.
This is effected by replacing the standard momentum operator with a Dunkl operator,
which in the $A_{N-1}$ case is defined as follows~\cite{dunkl},
\be
\label{dunkl}
\nabla_i=\partial_i-\sum_{j\neq i}\frac{g}{x_i-x_j}s_{ij}
\qquad\textrm{with}\qquad
[\nabla_i,\nabla_j]=0
\quad\textrm{and}\quad
[\nabla_i, x_j]=S_{ij}=
\begin{cases}
 -g s_{ij}&\text{for $ i\ne j$},\\
1+g\sum_{k\ne i} s_{ik} & \text{for $i= j$}.
\end{cases}
\ee
Here $s_{ij}$ is a $A_{N-1}$ Coxeter reflection operator, which acts as the permutation operator
exchanging the $i$th and $j$th coordinate:
\be
s_{ij}\psi(\dots, x_i,\dots, x_j, \dots) =\psi(\dots, x_j,\dots, x_i, \dots).
\ee

Formally, the Calogero interaction in the Hamiltonian is hidden in the nonlocal `connection' entering
the Dunkl-covariant Laplacian.
After a restriction to totally symmetric wave functions one obtains a local bosonic Hamiltonian
with a Calogero interaction potential. In other words, replacing partial derivatives by Dunkl operators in the
Hamiltonian of the $N$-dimensional harmonic oscillator, one gets the Calogero model in an oscillator potential.
Making the same substitution in the symmetry generators, we arrive at the constants of
motion of the Calogero-oscillator system. The picture is reminiscent of the nonlocal unitary transformation
mapping the Calogero particles to free ones~\cite{free}.

In our recent paper~\cite{CalCoul}, we have indicated that the spherical or hyperbolic extension of the
rational Calogero potential associated with an arbitrary Coxeter group is the only possible
superintegrable deformation of the $N$-dimensional oscillator and Coulomb systems.
The hidden symmetries of the quantum Calogero-oscillator system are well known~\cite{calogero-root,adler}.
Recently, explicit expressions for the constants of motion and the symmetry algebra of the quantum
Calogero-Coulomb model~\cite{khare} have also been revealed  within the Dunkl-operator approach in two~\cite{vinet}
and arbitrary~\cite{Runge} dimensions.

Lately, the same method has been applied to the integrable two-center Calogero-Coulomb and Calogero-Coulomb-Stark
systems~\cite{Calogero-Stark}.
It seems that any isotropic integrable system in $N$~dimensions can be Dunkl-deformed to add a Calogero-type
interaction respecting integrability.
Looking at the Coulomb system, it perfectly works for the angular momentum,
while for the Dunkl-extended Runge-Lenz vector we need to add some correction~\cite{Runge}.
The symmetry algebras of these Dunkl-deformed systems are nonlocal deformations of the initial ones.

We have also revealed also the superintegrability of the oscillator/Coulomb systems on the sphere~\cite{schroedinger,higgs}
in the presence of an extra Calogero potential \cite{CalCoul}.  Applying the matrix-model reduction, we have described
the symmetries of these systems in the classical case  \cite{classical}.
This method works  for the quantum systems too.
However, it forces to take into account the ordering of the individual entries in a matrix product,
making the calculations less transparent.

In the this paper we apply the exchange-operator approach to
the oscillator/Coulomb quantum models on the $N$-dimensional sphere with an additional Calogero potential.
There are obvious obstacles in this way, in particular:
\begin{itemize}
\item The Dunkl operators are not invariant even under the linear symmetry transformations,
      so we should first find their proper definition on the sphere.
\item We need to fix the operator ordering in the deformed quantities
      on curved spaces (including the sphere).
\end{itemize}

Our key point is the coordinate frame used in~\cite{CalCoul,classical}.
Namely, we parameterize the sphere by $N$~Cartesian coordinates $\bm{x}=(x_i)$ in the ambient Euclidean
space $\mathbb{R}^{N+1}\ni(x_0,\bm{x})$ defining the following metric:
\be
ds^2=d{\bm{x}}^2 +dx_{0}^2\vert_{x^2_0 + \bm{x}^2 = r^2_0}
=d\bm{x}^2+\frac{(\bm{x}\cdot d\bm{ x})^2}{r^2_0-\bm{ x}^2}
=h_{ij}dx_idx_j
\qquad\textrm{so that}\qquad  h_{ij}=\delta_{ij}-\frac{x_ix_j}{r^2_0}.
\label{metric}
\ee
First, we choose  the Dunkl operators on the sphere by using the same coordinate expressions
as in the flat case~\eqref{dunkl}.
Second, we impose an operator ordering for the hidden-symmetry generators and Hamiltonians of the
spherical Dunkl-oscillator and Dunkl-Coulomb systems.
Third, the replacement $\bm{\partial}\to \bm{\nabla}$ then leads to the correct expressions for all quantities.
Fourth, we calculate the symmetry algebras and find deformations of those of the familiar
spherical oscillator/Coulomb systems.

The paper is organized as follows.
In Section 2 the Hamiltonians of the Calogero-oscillator and Calogero-Coulomb systems on the sphere are formulated
in terms of Dunkl operators.
In Section 3 the symmetry generators of the spherical Calogero-oscillator system are constructed
and their algebra is evaluated.
In Section 4 we find the analog of the Runge-Lenz vector and compute the symmetry algebra for the
spherical Calogero-Coulomb system.

\section{General Consideration}
\noindent
According to the general prescription for quantum systems on the $N$-dimensional sphere
$S^N\hookrightarrow\mathbb{R}^{N+1}$,
the kinetic part of a Hamiltonian is given by the Laplace-Beltrami operator,
\be
\sum\limits_{i,j=1}^N \frac{1}{\sqrt{h}}\partial_i \left(\sqrt{h} h^{ij} \partial_j \right)
=
\bm{\partial}^2 - \frac{1}{r_0^2}\left(\bm{x}\cdot\bm{\partial}+N{-}1 \right)(\bm{x}\cdot\bm{\partial})
=\bm{\partial}^2-\frac{1}{4r_0^2}\left(\left\{\bm{x},\bm{\partial}\right\}^2-2\left\{\bm{x},\bm{\partial}\right\}
-N(N{-}2)\right) .
\label{lbop}
\ee
Here $h=\det h_{ij}={r_0^2}/(r^2_0-\bm{ x}^2) $ and $h^{ij}=\delta^{ij}- x^i x^j/r_0^2 $ are, respectively
the determinant and the inverse metric on the sphere~\eqref{metric}.
For the inclusion of the Calogero interaction, we replace
 the partial derivatives in the symmetrized version of~\eqref{lbop}  by the Dunkl
operators \cite{dunkl} and get the following nonlocal Hamiltonian on the sphere:
\be
\label{Hgen}
\begin{aligned}
\mathcal{\cal H}_\text{osc/Coul}
&= -\frac12 \left[ \bm{\nabla}^2
-\frac{1}{4r_0^2}\left(\left\{\bm{x},\bm{\nabla}\right\}^2-2\left\{\bm{x},\bm{\nabla}\right\}-N(N{-}2) \right)\right]
+ V_\text{osc/Coul}(x)
\\
&=-\frac12 \bm{\partial}^2
+\frac{1}{2r_0^2}\left(\bm{x}\cdot\bm{\partial}+N{-}1 \right)(\bm{x}\cdot\bm{\partial})
+\sum_{i< j}^N
\frac{g\,(g-s_{ij})}{(x_i-x_j)^2}+ V_\text{osc/Coul}(x),
\end{aligned}
\ee
where the $ V_\text{osc/Coul}$ is the oscillator/Coulomb potential on the sphere given by the expressions \cite{higgs}
\be
V_\text{osc}=\frac{r^2_0}{x^2_0}\frac{\omega^2x^2}{2}
\qquad\textrm{and}\qquad
V_\text{Coul}=-\frac{x_0}{r_0}\frac{\gamma}{x}
\qquad \textrm{with} \qquad
x=|\bm{x}|.
\label{potentials}
\ee
The second equation in \eqref{Hgen} follows from the identity
$\sum_i\{\nabla_i , x_i\}=\sum_i\{\partial_i,x_i\}$, which
is a direct consequence of the commutation relations~\eqref{dunkl} among Dunkl derivatives and coordinates.
The restriction on symmetric wave functions produces the spherical system with an additional Calogero potential:
\be
\label{H}
{ H}_\text{osc/Coul}= {\rm Res}\,  \mathcal{\cal H}_\text{osc/Coul} = -\frac12 \bm{\partial}^2
+\frac{1}{2r_0^2}\left(\bm{x}\cdot\bm{\partial}+N{-}1 \right)(\bm{x}\cdot\bm{\partial})
+\sum_{i< j}^N
\frac{g(g{-}1)}{(x_i-x_j)^2}+ V_\text{osc/Coul}(x).
\ee

The  generalized Hamiltonian \eqref{Hgen}  commutes with
the Dunkl angular momentum  inherited from the flat case~\cite{kuznetsov,feigin}:
\be
\label{Lij}
L_{ij} = x_i \nabla_j - x_j \nabla_i
\qquad\textrm{obeys}\qquad
\big[ L_{ij},\mathcal{\cal H}_\text{osc/Coul} \big] = 0.
\ee
The related algebra has recently been investigated in detail~\cite{fh}.
In particular, the deformed generators satisfy the  following  commutation relations,
\be
\label{comLL}
\left[ L_{ij} , L_{kl}  \right]=S_{jk} L_{il}+S_{il} L_{jk}-S_{ki} L_{jl}-S_{lj} L_{ik}
\ee
with the modified permutation operators $S_{ij}$ defined in~\eqref{dunkl}.

The corresponding symmetries of the restricted Hamiltonian \eqref{H}
are given by the symmetrized powers
\be
\label{Lk}
{\cal L}_{2k}=\sum_{i<j} L_{ij}^{2k}.
\ee
The first integral  is essentially the Casimir element of the Dunkl angular momentum algebra.
It is proportional to  the angular part~${\cal L}'_2$ of the generalized Calogero Hamiltonian \cite{fh},
\be
\label{L2'}
{\cal L}'_2 = {\cal L}_2 - S(S{-}N{+}2)
\quad\textrm{with}\quad
S=\sum_{i<j} S_{ij}
\qquad\textrm{so that}\qquad
\big[{\cal L}'_2, L_{ij}\big]=0=[S,s_{ij}].
\ee
The angular Calogero model ${\cal L}'_2$ has been studied quite thoroughly~\cite{sphCal,flp,sph-mat,fh}.
In particular, its spectrum and wave functions have been derived \cite{flp}, and the
classical \cite{sph-mat} and quantum \cite{fh} symmetry algebra have been investigated.

Finally note that, if we Dunkl deform the non-symmetrized version of the Laplace-Beltrami operator~\eqref{lbop},
our Hamiltonian \eqref{Hgen} will pick up an additional $S$-term since
\begin{align}
\sum_{i,j} \frac{1}{\sqrt{h}}\nabla_i \left(\sqrt{h} h^{ij} \nabla_j \right)
=\bm{\partial}^2
-\frac{1}{r_0^2}\left(\bm{x}\cdot\bm{\partial}+N{-}1 \right)(\bm{x}\cdot\bm{\partial})
-\sum_{i\ne j}^N \frac{g\,(g-s_{ij})}{(x_i-x_j)^2}
+\frac{1}{r_0^2} S(S{-}N{+}1).
\end{align}
Because the difference between the two versions reduces to a number on symmetric wave functions,
this is inconsequential, and we use the former one.

\section{Integrals of the Calogero-oscillator system on the sphere}
\noindent
Let us consider  the Dunkl representation~${\cal H}_\text{osc}$ of the quantum Calogero-oscillator system on the sphere,
given by the generalized Hamiltonian \eqref{Hgen}  with the potential $V_\text{osc}$ from \eqref{potentials},
which we will refer to as the `spherical Dunkl oscillator'.
We choose the following ansatz for the generators of its hidden symmetries:
\begin{align}
\label{Iij}
I_{ij}=\frac1{2r_0^2}\left\{x_0\nabla_i , x_0\nabla_j \right\} - \omega^2 r_0^2 \frac{x_i x_j}{x_0^2}
\qquad\textrm{so that}\qquad
[I_{ij}, {\cal H}_\text{osc}]=0.
\end{align}
It generalizes the well known Fradkin tensor for the
flat isotropic oscillator~\cite{fradkin} and its extension to the sphere~\cite{higgs}.

It is not hard to verify the commutation relations with the Dunkl angular momenta generators,
\be
\label{comLI}
\left[ L_{ij} , I_{kl}  \right]=I_{ik}S_{jl}+S_{jk}I_{il}-I_{jk}S_{il} -S_{ik}I_{jl} +\frac{1}{2r_0^2}\left[ L_{ij}, S_{kl} \right].
\ee
The commutations between the hidden-symmetry generators~\eqref{Iij} are more sophisticated,
\be
\label{comII}
\begin{split}
\left[ I_{ij} , I_{kl}  \right] =&-\frac{1}{r_0^2}\Big( I_{il}L_{jk}+ L_{jl} I_{ik} + I_{jk}L_{il}
    + L_{ik} I_{jl} \Big)-\frac{1}{2r_0^2}\Big( \left[ S_{ij}, I_{kl} \right]-\left[ S_{kl}, I_{ij} \right] \Big)
    \\
&+\left(\omega^2-\frac{1}{4r_0^4}\right)\Big(  S_{jl} L_{ik} + L_{il} S_{jk} + L_{jk} S_{il} +S_{ik} L_{jl}  \Big)
    +\omega^2 \left[S_{ij},S_{kl} \right].
\end{split}
\ee

Now we switch to the spherical Calogero-oscillator Hamiltonian~\eqref{H}
by restricting to the subspace of symmetric wave functions.
Evidently, any permutation-invariant combination
of products of the elements \eqref{Lij} and \eqref{Iij}
will produce an integral of motion. In particular, the constants of motion
 ${\cal L}_{2k}$
\eqref{Lk} of the generalized angular Calogero Hamiltonian ${\cal L}'_2$ are preserved here too.
Furthermore, the deformed Fradkin tensor \eqref{Iij}
provides two series of symmetric powers which are of $2k$-th order in momenta:
\be
\label{Ik}
{\cal I}^{(1)}_k = \sum_{i} I_{ii}^k,
\qquad
{\cal I}^{(2)}_k = \sum_{i,j} I_{ij}^k.
\ee
The invariants constructed so far already include a full set of $2N{-}1$ functionally independent
integrals of the Calogero-oscillator Hamiltonian on the sphere.
The generalized Hamiltonian may be expressed in terms of these,
\be
{\cal H}_\text{osc}=-\frac12{\cal I}^{(1)}_1 -\frac{{\cal L}'_2+S}{2r_0^2}.
\ee
The first integral from the second family in~\eqref{Ik} depends only on the center-of-mass coordinates,
\be
{\cal I}^{(2)}_1 = \frac{x_0^2}{r_0^2} D^2-\frac{1}{r_0^2}XD - \frac{\omega^2r_0^2}{x_0^2}X^2
\qquad
\text{with}
\qquad
X=\sum_i x_i,
\quad
D=\sum_i \partial_i.
\ee

\bigskip

\noindent
\textbf{Flat-space limit.}
In the limit $r_0\to \infty$, the Dunkl angular momentum operators $L_{ij}$
with their commutators remain unchanged.
The model is reduced to the conventional  Calogero-oscillator system, which
can be expressed in terms  of deformed
creation-annihilation operators~\cite{brink}:
\begin{gather}
{\cal H}_\text{osc/flat}=-\frac{1}{2} \bm{\nabla}^2+ \frac{\omega^2 }{2} \bm{x}^2
=\frac\omega2\sum_i (a_i^+a_i + a_i a^+_i)
\\
\label{a-pm}
\textrm{where} \qquad
a_i=\frac{\omega x_i+ \nabla_i}{\sqrt{2\omega}},
\qquad
a_i^+=\frac{\omega x_i- \nabla_i}{\sqrt{2\omega}}.
\end{gather}
They obey the Dunkl-operator commutations~\eqref{dunkl}
after the replacement $x_i\to a^+_i$ and $\nabla_i\to a_i$.

In the flat-space limit,
the symmetry generators \eqref{Lij} and \eqref{Iij} simplify to
\begin{align}
L_{ij}=a^+_ia_j-a^+_ja_i
\qquad\textrm{and}\qquad
I_{ij}=-\omega(a^+_ia_j + a^+_ja_i + S_{ij}),
\end{align}
and their algebra \eqref{comLL}, \eqref{comLI} and \eqref{comII} reduces to the deformed $u(N)$ algebra
investigated in detail in~\cite{fh}.
In particular, the crossing relations
\be
\label{cros}
E_{ij} E_{kl}- E_{il} E_{kj}= E_{il}S_{kj}-E_{ij}S_{kl}
\ee
among the generators $E_{ij}=a_i^+a_j$ imply the commutation relations
\be
\label{comEE}
[E_{ij}, E_{kl}]= E_{il}S_{jk} - S_{il} E_{kj} + [S_{kl}, E_{ij}].
\ee
The latter agrees with the relations obtained in the flat-space limit from~\eqref{comLI} and \eqref{comII},
\begin{gather}
\left[ L_{ij} , I_{kl}  \right]=I_{ik}S_{jl}+S_{jk}I_{il}-I_{jk}S_{il} -S_{ik}I_{jl}, \\
\left[ I_{ij} , I_{kl}  \right]= \omega^2 \big(  S_{jl} L_{ik} + L_{il} S_{jk} + L_{jk} S_{il} +S_{ik} L_{jl}  + \left[S_{ij},S_{kl} \right]\big).
\end{gather}
Although the deformations of the Cartan algebra elements $E_{ii}$ (or of $I_{ii}=-2E_{ii}-S_{ii}$) do not commute,
the related symmetric polynomials mutually commute as was proven in~\cite{poly92},
\be
\big[{\cal I}^{(1)}_i, {\cal I}^{(1)}_j\big]=0.
\ee
They form a set of Liouville integrals of the Calogero-oscillator system.

\section{Integrals of the Calogero-Coulomb system on the sphere}
\noindent
The Dunkl representation of the quantum Calogero-oscillator system on the sphere is  given by the generalized Hamiltonian
\eqref{Hgen}  with the Coulomb potential $V_\text{Coul}$~\eqref{potentials}, and we shall
refer to it as the spherical Dunkl-Coulomb system.
Knowing the deformed Runge-Lenz vector of the flat Dunkl-Coulomb system~\cite{Runge},
we choose the following ansatz for its extension to the sphere,
\be
\label{Ai}
A_{i}=-\frac{x_0}{2r_0}\sum_{j=1}^N\left\{ L_{ij} , \nabla_{j}  \right\}
    +\frac{x_0}{2r_0}\left[ \,\nabla_{i} ,S\, \right]-\gamma \frac{x_i}{x}
\qquad\textrm{which indeed obeys}\qquad
\left[ A_{i}, {\cal H}_\text{Coul}\right]=0.
\ee
After some simple algebra the above expression can be recast as
\be
\label{Ai-2}
A_i=\frac{x_0}{r_0}\Big(\bm{x}\cdot\bm{\partial}+\frac {N{-}1}2\Big)\nabla_i
- x_i \Big(\frac{x_0}{r_0}\bm{\nabla}^2+\frac\gamma x\Big).
\ee
The commutation relations of the Dunkl angular momentum with the deformed Runge-Lenz vector
remain as they are in the flat case~\cite{Runge},
\be
\label{comLA}
\left[ L_{ij} , A_{k}  \right]=-S_{ik}A_{j} +S_{jk}A_{i}.
\ee
The components of the deformed Runge-Lenz vector commute as follows,
\be
\label{comAA}
\left[ A_{i} , A_{j}  \right]=-2{\cal H}'L_{ij},
\ee
where we have introduced the  operator
\be
\label{H'}
{\cal H}'={\cal H}_\text{Coul}+\frac{1}{r_0^2}\left( {\cal L}'_2 - \frac{(N{-}3)^2}{8}\right)
\qquad\textrm{which still obeys}\qquad
\big[{\cal H}', L_{ij}\big]=0,
\ee
but does not commute with the deformed Runge-Lenz vector any more.

In order to find the integrals of motion of the spherical Calogero-Coulomb model~\eqref{H},
obtained by restricting the generalized Hamiltonian~\eqref{Hgen} to totally symmetric
wave functions, we have to combine the constructed invariants into symmetric polynomials
as was described already for the oscillator case. They are given by the ${\cal L}_k$ \eqref{Lk}
and the following $2k$-th order (in momentum) invariants,
\be
\label{Ak}
{\cal A}_k=\sum_i A_i^k.
\ee
The first member of this family is deduced immediately from~\eqref{Ai-2}
and depends only on the center-of-mass degree of freedom:
\be
\label{A1}
{\cal A}_1 =\frac{x_0}{r_0}  \Big(\bm{x}\cdot\bm{\partial} + \frac{N-1}{2}\Big)D  - X \Big(\frac{x_0}{r_0}\bm{\nabla}^2 +\frac\gamma x\Big).
\ee
The second member   is just the square of the deformed Runge-Lenz vector,
${\cal A}_2=\bm{A}^2$, and depends on the simpler integrals.
As a consequence of the commutation relations \eqref{comLA}, \eqref{comAA} and \eqref{H'},
a simple modification of it commutes with the angular momentum:
\be
{\cal A}'_2={\cal A}_2+2{\cal H}'S
\qquad\textrm{obeys}\qquad
\big[{\cal A}'_2,L_{ij}\big]=0.
\ee
Thus one can expect that it can be expressed in terms of the generalized 
angular Calogero~\eqref{L2'} and
 Dunkl-Coulomb~\eqref{Hgen}  Hamiltonians.
In fact, the explicit relation between these three quantities is given by
\be
{\cal A}'_2 =\gamma^2 -2{\cal H}'\left( {\cal L}'_2 - \frac{(N-1)^2}{4}\right)
+\frac{1}{r_0^2}\left( {\cal L}'_2 - \frac{(N-1)(N-3)}{4}\right)^2.
\ee

\bigskip

\noindent
\textbf{Flat-space limit}.
In the limit  $r_0\to\infty$ we arrive at the Calogero-Coulomb model studied in detail recently in~\cite{Runge}.
Some of our expressions then simplify. In particular, \eqref{H'} reduces to $\cal{H}'={\cal H}_\text{Coul}$.
The integrals of motion and their algebra are mapped to those derived there for the flat case.

\acknowledgments
\noindent
T.H.\ and A.N.\ were partially supported by the Armenian
State Committee of Science Grants No.~15RF-039 and No.~15T-1C367 and by Grant No.~mathph-4220 of the
Armenian National Science and Education Fund based in New York (ANSEF). The work of T.H.\ and A.N.\
was done within the ICTP programs NET68 and OEA-AC-100 and within the program of Regional Training Networks
on Theoretical Physics by VolkswagenStiftung contract nr.~86~260.
The work of F.C.\ is supported by the Alexander von Humboldt Foundation under grant CHL 1153844 STP.

\end{document}